\documentclass[manuscript]{aastex}
\usepackage{graphicx}
\usepackage{natbib}

\shorttitle{Effects of Magnetic field on 6173 and 6302 nm lines}
\shortauthors{Criscuoli et al.}

\begin{document}

   \title {Effects of unresolved magnetic field\\ on Fe I 617.3 and 630.2 nm line shapes
    }
   \author{S.~Criscuoli\altaffilmark{1,2},  I.~Ermolli\altaffilmark{1},  H.~Uitenbroek\altaffilmark{2} and F.~Giorgi\altaffilmark{1}}

    \altaffiltext{1}{INAF-Osservatorio Astronomico di Roma, Via Frascati 33, 00040 Monte Porzio Catone, Italy}
\altaffiltext{2}{ National Solar Observatory, Sacramento Peak, P.O. Box 62, Sunpsot, NM 88349, USA}

   \date{}

  \begin{abstract}
The
contribution of the quiet Sun to solar irradiance variability, due either to changes of the solar effective temperature or to the presence of unresolved magnetic field,  is still poorly understood. In this study we investigate spectral line diagnostics that are sensitive to both temperature variations and the presence of small scale unresolved magnetic features in these areas of the solar atmosphere.
Specifically we study the dependence on the magnetic flux density of three parameters describing the shape of two magnetically sensitive \ion{Fe}{1} lines, at 630.2 nm and 617.3 nm, namely the line core intensity (IC), full width at half maximum (FWHM), and the equivalent width (EQW). To this aim we analyze observations of active region NOAA 11172, acquired with IBIS at the Dunn Solar Telescope, as well as results from numerical synthesis. Our results show that IC is sensitive to both temperature and magnetic flux density variations, FWHM is mostly affected by magnetic field changes, and EQW is mostly sensitive to temperature. Variations of a few percent of the measured line parameters are found in observational data that was spatially degraded to represent quiet-Sun, disk-centre, medium resolution observations. It is therefore possible to disentangle magnetic from pure thermodynamic effects by comparison of temporal variations of the EQW and the FWHM of either the two \ion{Fe}{1} lines.  
\end{abstract}


\maketitle

\section{Introduction}
It is well established that both the total and spectral Solar Irradiance (SI, henceforth) are modulated by the emergence and evolution of the magnetic field in the solar atmosphere. In particular, SI variations measured on time scales from days to several solar rotations can be explained by accounting for the radiative effect of the presence of 
magnetic flux in sunspots, faculae, and network regions \citep[e.g.][]{domingo2009}. However, there is still uncertainty in measurements of SI 
variations on time scales of the activity cycle and longer \citep[e.g.][]{harder2009,frohlich2011}, as well as in the determination of individual contributions of 
different magnetic features to the measured variations \citep[e.g.][]{krivova2011,fontenla2012}. The role of the quiet Sun, that is those areas on the solar surface away from magnetically active and network regions, the so called Inter Network regions, is particularly debated. For instance,  long-term SI variations unexplained by some recent models have been attributed either to a 0.2 K global change in the Sun's effective temperature \citep{frohlich2011} or to quiet Sun variations associated with small-scale magnetic field changes during the activity cycle \citep{fontenla2012}.

Two main difficulties hinder the verification of these hypotheses. First, investigations of temperature variations would be achievable only by measurements of  absolute radiometric contribution of the quiet Sun, and such measurements are not available yet. Second,
 the  small-size magnetic features (100 km) that are seen permeating the whole solar surface in high-resolution observations \citep[and references therein]{bonet2012},  are unresolved in full-disk observations usually employed for SI reconstructions \citep{domingo2009}, so that their contribution to SI variations is hidden. 
Note that the origin of this so called quiet-Sun magnetic field is also controversial, as it is not clear yet whether and to what extent,  it results from the decay of active regions, turbulent dynamo effects or both \citep[e.g.][]{lites2011, lopezariste2012}.  Variations of quiet Sun magnetic flux over the cycle would indicate a global dynamo origin of the field; nevertheless, most of such measurements seem to indicate no temporal variations of quiet Sun field \citep[][and references therein]{sanchezalmeida2011} thus suggesting a local dynamo origin. Whether this result is due to lack of sensitivity of instrumentation and the observational techniques employed, or to a true constancy of the field over the cycle is still an open issue. 

The presence of magnetic field influences the emitted radiation in magnetically sensitive lines directly, by the Zeeman effect, and indirectly, by changing the thermodynamic properties of the plasma \citep[e.g.][]{fabbian2010}. Variations of line properties have been therefore employed to study the properties of quiet Sun. For instance, \citet{stenfloandlindegren1977} %
derived, by the investigation of  the subtle broadening of a large sample of \ion{Fe}{1} lines, an upper limit for the magnetic field amplitude of quiet Sun network  of 1000-2000 Gauss and an upper limit of 90 Gauss for intranetwork regions. Moreover, line parameters of a set of Fraunhofer lines were 
observed from 1974 to 2006 \citep{livingston2007}, with the aim of inferring long-term variations of the photospheric temperature. Analysis of these data lead, however, to inconclusive results \citep{penza2004,livingston2007}.  

In this study we investigate the effect of magnetic features on the shape of two \ion{Fe}{1} magnetically sensitive photospheric lines, the  630.2 and 617.3 nm, by analysing high-resolution spectro-polarimetric data of NOAA \textbf{11172} observed on March 17th, 2011 at the Dunn Solar Telescope with the Interferometric Bidimensional Spectrometer \citep[IBIS][]{cavallini2006,reardon2008}. Qualitative comparison with results from numerical synthesis has been performed to disentangle between the direct and indirect effects of magnetic field on the measured line shapes  variations.  We then investigate the effects of unresolved magnetic features on the shapes of the two lines by analysing IBIS data spatially degraded to mimic medium resolution observations. 
The \ion{Fe}{1} 630.2 and 617.3 nm lines were chosen because they have been frequently employed for photospheric magnetic field studies. Moreover, data at these spectral ranges are currently acquired with several instruments, like the HINODE/SOT \citep{tsuneta2008}, SOLIS/VSM \citep{keller2003}, and SDO/HMI \citep{scherrer2012}. 

The article is organized as follows: in Sec. 2 we describe the observations, the data reduction and the numerical synthesis; in Sec. 3 we discuss the line formation properties of the two lines; in Sec. 4 we 
present the results derived from analyses of observational data; in Sec. 5 we describe results obtained from synthesis; in Sec. 6 we investigate the effects of unresolved magnetic features; in Sec. 7 we discuss the results and draw our conclusions. 


%
\section{Data}
We investigated the behavior of the two \ion{Fe}{1} lines at 617.3 and 630.2 nm in data acquired at the NSO Dunn Solar Telescope with IBIS. In order to interpret the results derived from observations in terms of properties of the observed solar plasma, we also synthetized the two lines in semi-empirical atmospheric models representative of quiet Sun and magnetic regions.
\begin{figure*} 
\epsscale{.75}
\includegraphics[width=6cm, trim=0cm 0cm 7cm 0cm] {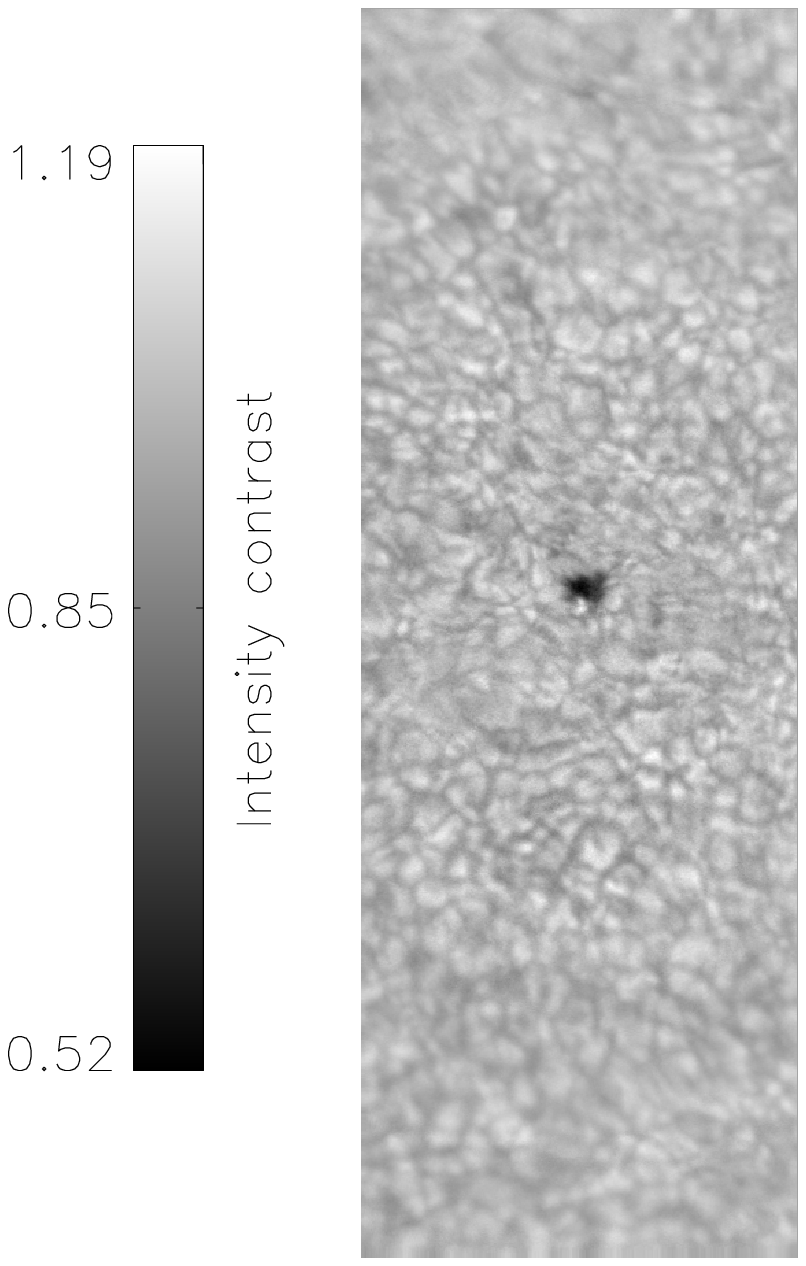}
\includegraphics[width=6cm, trim=0cm 0cm 7cm 0cm]{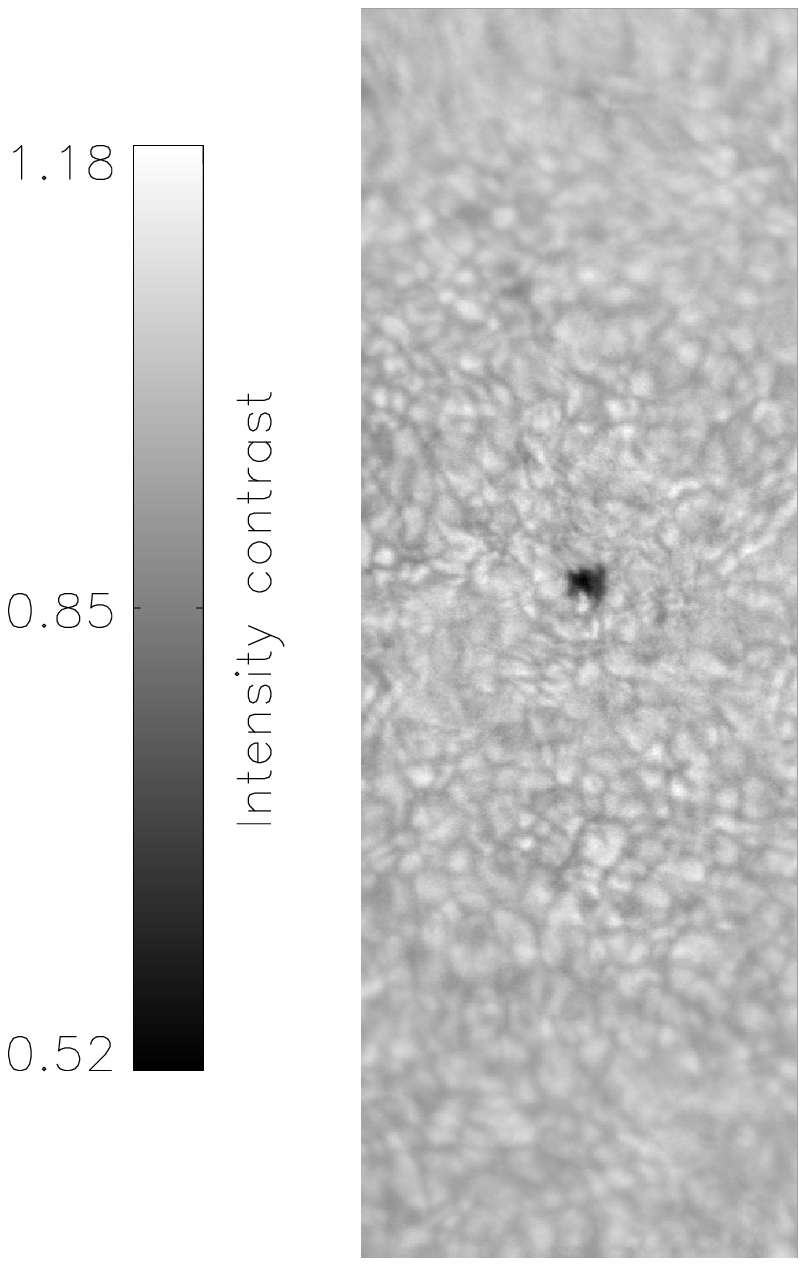}\\
\includegraphics[width=6cm, trim=0cm 0cm 7cm 0cm]{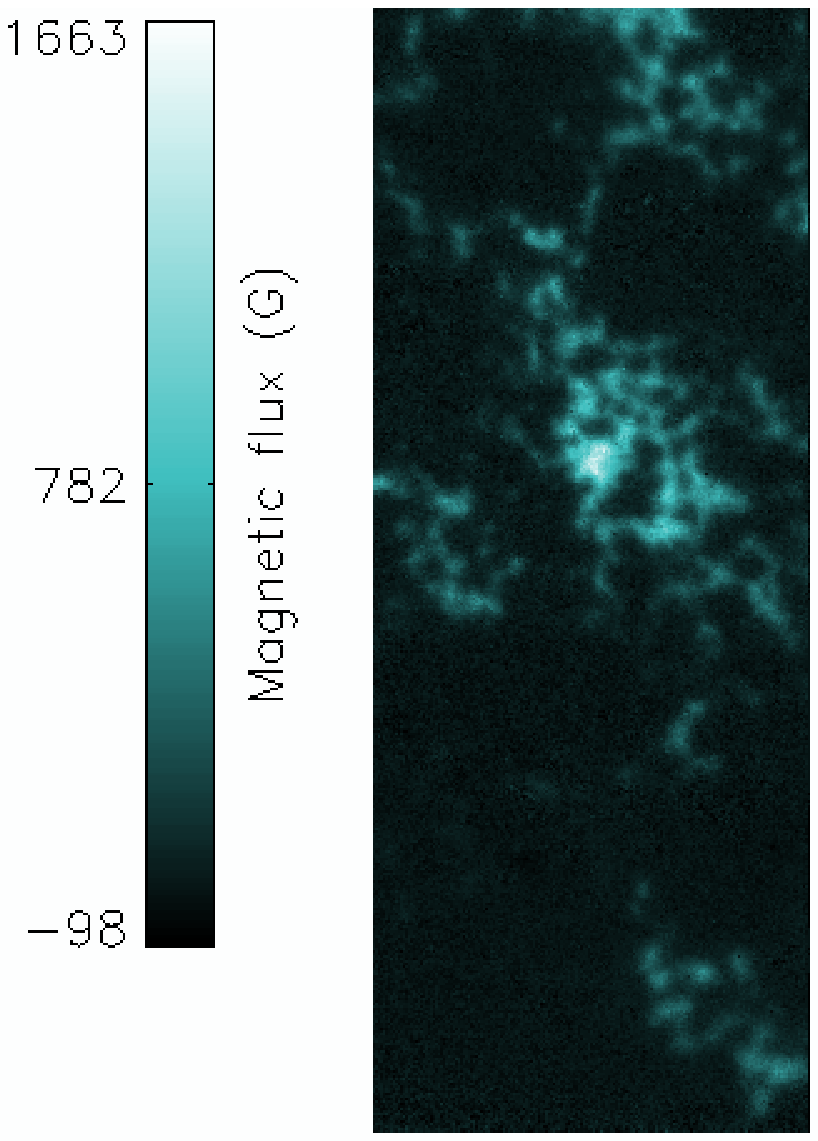}
\includegraphics[width=6cm, trim=0cm 0cm 7cm 0cm]{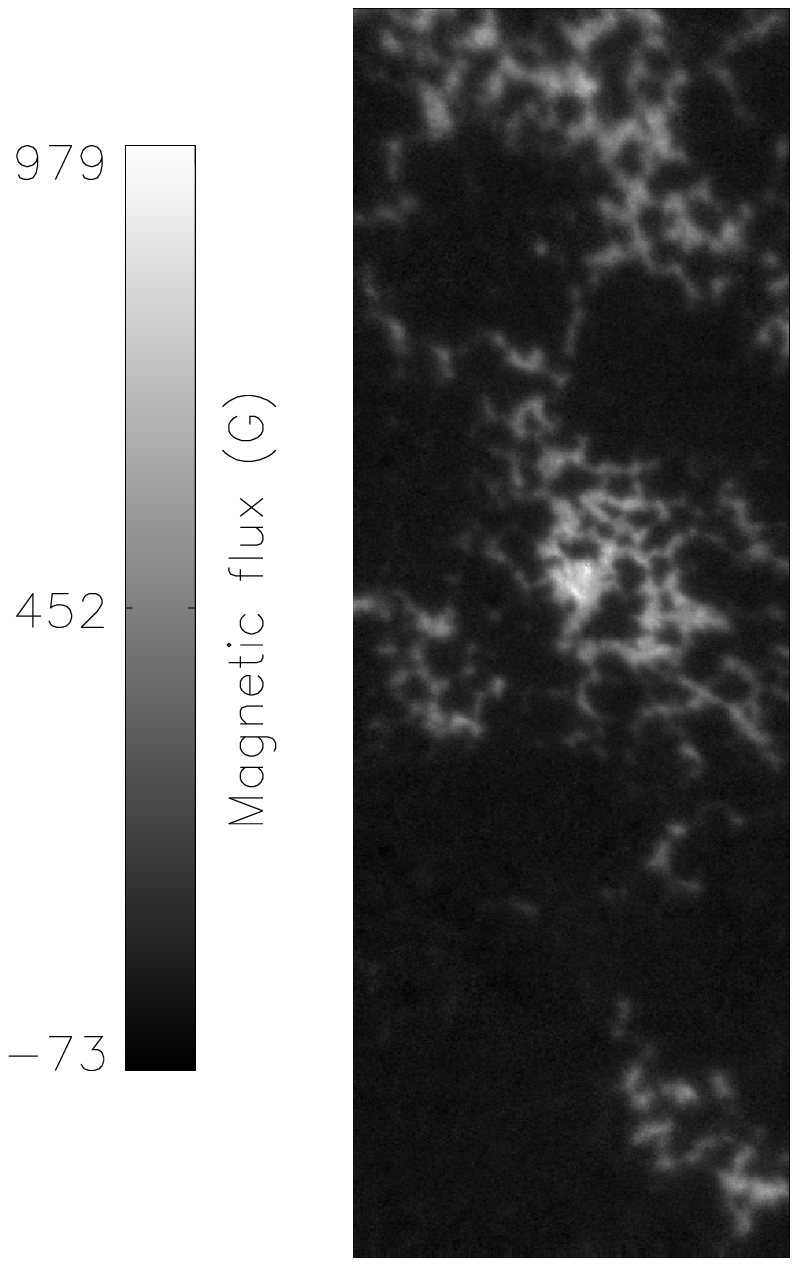}\\
\caption{Examples of analyzed IBIS data. Top: 617.3 nm (left) and 630.2 nm (right) nearby continua. Bottom: corresponding magnetic flux density maps derived with the COG method. The FOV is approximately 40 $\times$ 90 arcsec. }  

\label{dataexample} 
\end{figure*} 
\subsection{Observations and data analysis}
\label{sec:data}
The IBIS data set employed for this study was taken on March 17th 2011 from 14:19 UT to 15:19 UT. It consists of 60 spectro-polarimetric scans of the 
two lines, each sampled at 22 spectral positions; the nearby $0_{2}$ 630.28 nm Telluric line was also sampled at 6 
spectral positions and served for line position calibration purposes.  A full scan of the two lines took $\approx$ 60 seconds, so that each line was scanned in $\approx$ 30 seconds. This temporal cadence is short with respect to typical evolution time scales of solar photospheric phenomena, so that the integrity of each line during a scan can be assumed to be preserved. The Field of View (FOV) was approximately 45$\times$90  arcsec and included a facular region and a small 
pore located at  [16.5N,10.3E] in between the leading and following parts of NOAA 11172.  White Light (621.3 nm) frames imaging the same FOV were acquired 
contemporaneously to spectro-polarimetric data. The spatial sampling of IBIS and WL frames was 0.17\arcsec pixel$^{-1}$. 

The High Order Adaptive Optics System of the DST \citep{rimmele2004} was tracking on a small pore in the FOV during the acquisition, thus improving image quality and stability. The data set was reduced and compensated for instrumental blue-shift, and for instrument and telescope
polarization  by standard procedures \citep[i.e.][]{cauzzi2008, viticchie2009, judge2010}. Residual aberrations introduced by the atmospheric seeing were estimated  by applying Multi Frame Blind deconvolution technique  \citep{vannoort2006} to WL images and were employed to destretch the spectro-polarimetric images.  The residual crosstalk between the Stokes parameters and the rms noise level were estimated to be $2\cdot 10^{-3}$ and of the order of $10^{-3}$ the intensity in the nearby continuum, respectively.
An infrared blocker in the optical path reduced the spectral stray-light contamination, so that its residual, estimated as described in \citet{criscuoli2011}
is below 8 $\%$ and 4 $\%$ for 617.3 nm and 630.2 nm, respectively. 
For the purpose of this study, we analyzed a subsample of 10 scans having the highest rms-contrast in WL images. 
Figure \ref{dataexample} shows an example of the analysed data. 

We estimated the magnetic flux density along the Line-of-Sight in the FOV by applying the Center Of Gravity method \citep[COG;][]{rees1979} 
to the measured right- and left- circularly polarized signals. \citet{uitenbroek2003} 
showed that the COG provides good estimates of the magnetic flux
at the formation height of typical photospheric line core for fields lower than 1.5 kGauss. To estimate uncertainties of magnetic flux density values derived with the COG we applied standard error propagation to the COG formula. We found that the uncertainty is on average equal to 8 Gauss for flux density values smaller than 150 Gauss; it increases monotonically for larger flux density values, although the relative uncertainty is approximately constant and equal to 3$\%$.  The scatter of values increases with the decrease of the estimated flux. The average value of the uncertainties computed over the whole FOV is 9 Gauss, with a standard deviation of 6 Gauss. Note, however, that the finite polarimetric sensitivity of IBIS sets a lower limit of confidence on magnetic flux density estimates. To estimate this limit we computed the average absolute flux over those pixels in the FOV whose total polarization signal \footnote{The total polarization signal is defined as $\int{\sqrt{{U(\lambda)}^2+{V(\lambda)}^2+{Q(\lambda)}^2}d\lambda}/I_c$, where $I_c$ is the continuum intensty of Stokes I, and U, V and Q are the corresponding Stokes parameter. is lower than 3 $m\AA$ and found a value of 20 Gauss with a standard deviation of 10 Gauss. We therefore defined as quiet-Sun all those pixels whose absolute magnetic flux density is smaller that 30 Gauss (that is, below one standard deviation from the mean). }

Next, in order to characterize the shapes of the lines, we computed the core intensity (IC, hereafter), the equivalent width (EQW, hereafter), and the full width at half maximum (FWHM, hereafter) of the two lines at each image pixel of the FOV. 
In particular, the IC and FWHM were computed by fitting the line sampling with a Gaussian function.  
We computed the three line shape parameters only for those pixels in the FOV with a magnetic flux  below 1000 Gauss,thus excluding the small pore and the area around it from the analysis, but including network and facular elements. This allowed us to investigate the dependence of line shape parameters on magnetic flux density variations.

\subsection{Spectral synthesis}
\label{data:synthesis}
We performed synthesis of the two \ion{Fe}{1} line profiles in various one-dimensional semi-empirical static models representative of various thermodynamic conditions of the plasma and for different values of magnetic field strengths. In particular, we employed FAL-C, FAL-E, FAL-H and FAL-P  \citep{fontenla1999} models, representative of quiet Sun, network, facular and bright facular regions, respectively. The spectral synthesis was performed in local thermal equilibrium with the RH
code \citep{uitenbroek2002,uitenbroek2003}.  The Land\'{e} factor is 2.5 for both lines; the other atomic data adopted for the synthesis are summarized in Table \ref{tabellaid}. The three line parameters IC, FWHM and EQW were then computed for emergent spectra 
calculated at vertical line of sight.

\begin{table*}
\begin{center}
\begin{tabular}{ccc}
\tableline
Line (nm)  & $\log{gf}$ & $\chi$ (eV) \\
\tableline
\tableline

617.3 	  &  -2.88  & 2.22 \\
630.2 	  &  -1.24 & 3.68 \\
\end{tabular}
\end{center}
\caption{\label{tabellaid} Lines Atomic Data}
\end{table*}
\section{Response Functions}

\textbf{In order to discuss the results obtained from the two \ion{Fe}{1} lines we first investigated their formation heights. To this aim we analysed results derived from spectral synthesis of the two lines through FAL-C and FAL-P models, the latter with uniform magnetic field of 1000 Gauss}.
 We then computed the Response Function (RF, hereafter) of IC of both lines to perturbations of either the temperature (T\_RF) 
or magnetic field (B\_RF) on both atmosphere models. 

\begin{figure} 
\epsscale{.9}
\plotone{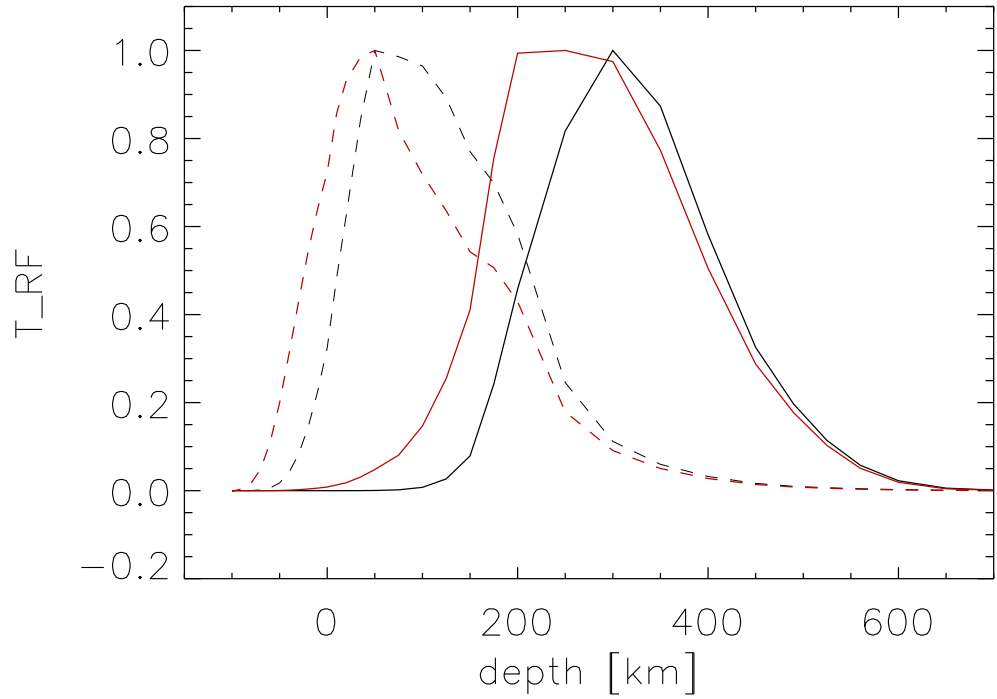}
\caption{Temperature Response Functions of line core intensity obtained in   FAL-C (continuous) and FAL-P (dashed) for the 630.2 nm (black) and the 617.3 nm (red) lines. Response Functions are normalized to their maxima. Optical depth unity at 500 nm sets the height reference value.\label{response_T} }
\end{figure} 

Figure \ref{response_T} shows the T\_RF of IC for both lines and models. 
We find that the 617.3 nm line core in FAL-C forms at a wide range of photospheric heights, from 0 to 600 km, with a peak at 200-300 km above
optical depth unity at 500 nm \citep[in agreement with][]{Norton2006}. 
The core of the 630.2 nm line instead forms in a slightly narrower range of photopsheric heights, from 180 to 600 km, and peaks slightly higher in the 
atmosphere at approximately 300 km. Figure \ref{response_T} also shows that the core of the two lines forms in the deepest layers of the  photosphere in magnetic 
features \citep[in agreement with][] {khomenkocollados2007}. This is the result of the lower temperature of FAL-P with respect to FAL-C at heights below optical depth unity in the latter (Fontenla et al. 1999), 
which is also found in computations of magnetic flux tube models \citep[e.g.][]{pizzo1993, steiner2005} and full 3-D Magneto Hydro Dynamic simulations \citep[e.g.][]{carlsson2004}. Both computed $RF_T$ peak at 100 km, with contribution from higher photospheric layers for the 630.2 nm line and from layers below optical depth unity for the 617.3 nm line. 
The bumps of both T\_RF at approximately 200 km reflect the change of the temperature gradient with height in FAL-P model. 

Figure \ref{response_B} shows the B\_RF of IC for both lines in the FAL-P model. The two  B\_RF are quite similar, as both peak at approximately 190 km, with the  B\_RF of the 617.3 nm line having again a slightly larger contribution 
from the deepest layers than the one calculated for the 630.2 nm line.       

The results derived from the synthesis indicate therefore that both lines form in similar regions of the solar atmosphere, with the 630.2 nm line sampling slightly 
higher atmospheric heights than the 617.3 nm line.

\section{Results from observations}
\label{measure}
In this section we investigate the dependence of the three line shape parameters on magnetic flux density. 

Note that we found that the flux density values 
estimated from the 630.2 nm data  are  on average lower than those obtained from the 617.3 nm set, and that the discrepancy increases with the magnetic flux density, with a maximum of approximately 40\% at 1000 Gauss. Because, as shown in \S 3,  the two lines form in similar regions of the solar atmosphere,  this discrepancy
is unlikely to reflect a stratification of the magnetic field with height. To verify this hypothesis we applied the COG to synthetic spectra of the two lines computed in the FAL-P model with a stratified magnetic field and obtained magnetic field estimates in agreement within 4\%.  We conclude that the discrepancy that we found in our data is most likely due to the Telluric blending affecting the 630.2 nm line. 
In the following, we therefore refer to flux density estimates derived from the 617.3 nm data only.

Results from our measurements are presented in Fig. \ref{shapes}. Dot symbols show the dependence of IC, EQW, and FWHM on the magnetic flux density for both lines, while asterisk symbols indicate the average value of the same line parameters computed over bins of magnetic flux. Note that the bins were increasingly larger in order to guarantee sufficient statistics. We found that the IC and FWHM increase with the magnetic flux for both lines, whereas the EQW shows the opposite trend. All three line shape parameters show saturation. For both IC and EQW this saturation occurs at approximately 300 Gauss for both lines, while  the FWHM saturates at flux density values above 700 Gauss and 500 Gauss for the 617.3 nm and the 630.2 nm line, respectively.
Note that the maximum relative variations with magnetic flux density (i.e. relative difference between values obtained at the largest flux and those obtained for quiet regions) are approximately 40\% for IC for both lines, 55\% and 36\% for 
FWHM in the case of 617.3 nm and 630.2 nm respectively, and -15\% and -8\% for EQW, respectively.

The measurements present large scatter, especially at the lowest flux density values, even though the standard deviations  of the values distributions in each bin (error bars in the plots) are rather small. The observed scatter of values reflects the sensitivity of line shape parameters to temperature and velocity variations of the plasma. The larger the magnetic flux, the more the temperature and velocity fluctuations of convective plasma motions are inhibited \citep[e.g.][]{ishikawa2007,romano2012}, so that the dispersion of the line shape parameters
becomes smaller and smaller with the increase of the magnetic flux density. 
Note that the dispersion for the 617.3 nm line is larger than that obtained for the 630.2 nm data, especially when comparing IC. This is due to the lower excitation potential of the 617.3 nm line, which makes the core intensity of this line more sensitive to temperature and velocity fluctuations \citep[][p. 285-287]{gray1992}. 

The IC in particular depends on both temperature and magnetic field in a manner that is difficult to disentangle. Results from numerical models and simulations of magnetic flux tubes show in fact that the photospheric temperature within a tube rises following an increase of magnetic field, due to the intensification of the lateral heating effects from the surrounding plasma \citep[e.g.][]{pizzo1993}. \textbf{As also qualitatively shown by results obtained from numerical synthesis  reported in Sec. \ref{appenidx:numsin}}, an increase in either of these quantities, the field strength, via the Zeeman effect, or the temperature, via irradiation, leads to an increase of IC, so that it is difficult to discern between the two effects.

The trend with the magnetic flux density of FWHM found for both lines suggests instead that this line shape parameter is more sensitive to magnetic field variations and velocity effects than to temperature. In fact, from the rise of temperature within flux tubes mentioned above, we would expect the FWHM to decrease with the magnetic flux. This decrease is compensated on one
 hand by the fact that magnetic flux concentrations tend to be located in or nearby downflow regions, which are characterized by shallower and broader profiles, and on the other hand, to the increase of the splitting of the lines with the magnetic flux. This latter effect becomes more and more important with the increase of the field (and therefore of the flux), due to the inhibition of convective flows mentioned above. This interpretation is confirmed by results obtained from numerical synthesis of line profiles discussed in Sec. \ref{appenidx:numsin}.

We also note that at low magnetic flux density values, the FWHM of 617.3 nm line increases slower with the magnetic flux density than for the 630.2 nm.  This indicates that for the former line the effect of temperature counteracting Zeeman broadening is larger than for the 630.2 nm.
Note that the maximum difference between the FWHM at saturation and the FWHM of average quiet Sun agree within the errors for the two lines (i.e. is approximately 0.06 \AA), as expected from Zeeman splitting theory for lines with equal Land\'e factor. 

Finally, the dependence of EQW on the magnetic flux reflects variations of both the width and depth of the studied lines. The rapid saturation of EQW with the magnetic flux is due to the fact that the line core intensity increases more rapidly than the FWHM for a given increase of magnetic flux. \textbf{ As formally shown in the Appendix \ref{SecA1}, and confirmed by results obtained from numerical synthesis shown in Sec. \ref{appenidx:numsin},} the EQW of the two iron lines is almost independent of the magnetic field and is mostly sensitive to temperature variations. Figure~\ref{shapes} therefore suggests that the magnetic features in our data are characterized by similar temperatures. 

\section{Results from numerical synthesis}
\label{appenidx:numsin}
In previous section we have analysed the dependence of the three shape parameters of the two iron lines on the magnetic flux density. As already noticed, this dependence is not explicit, because of the indirect effects of the magnetic field on other physical parameters. In particular, the indirect effect on the temperature stratification highly affects the shape of a line. In order to separately investigate the effects of magnetic field and temperature on line shape parameters we have therefore analysed results obtained from numerical synthesis through various one dimensional models, as described in Sec. \ref{data:synthesis}. Results are presented in this section.

To investigate the effects of temperature on the shapes of line profiles we have compared results obtained from the various FAL models, which are characterized by different temperature stratifications. Plots in Fig. \ref{EQWvsTsim} show the relative differences of the line parameters computed for FAL-E, FAL-H and FAL-P models with respect to line parameter obtained with FAL-C model.
Since the three models are characterized by increasing temperature at the formation heights of the two lines, the plots show that the increase of temperature causes  IC to increase and the FWHM and EQW to decrease. While for IC and EQW we found changes up to $50\%$ and more, the FWHM changes do not exceed $10\%$.

To investigate the effects of the magnetic field, we considered line profiles obtained with FAL-C model in the case of fields constant with geometric height and whose strength varied from 50 to 1000 Gauss. 
Results reported in Fig. \ref{EQWvsBsim}  show the relative difference between line parameters obtained for the different field strength values with respect to those obtained in the case of no magnetic field. We note that the increase of the magnetic field strength  leads to a large increase of IC and FWHM (more than $100\%$ in the case of IC), especially at the largest magnetic strength values, while EQW stays fairly constant (variations are smaller than $1\%$ for both lines), thus confirming the analytical derivation presented in Appendix \ref{SecA1}. 

Both plots show that the two lines are differently affected by changes of plasma physical properties, mainly because of their different atomic parameters. The plot in Fig. \ref{response_T} shows for instance that the difference of the shapes of the T\_RFs obtained for FAL-C and FAL-P models is larger for the 617.3 nm line than for the 630.2 nm line. In particular, while the T\_RFs of the 630.2 nm line obtained with the two models appear mostly shifted in depth respect to each other, the two T\_RFs obtained for the 617.3 nm line also show different shapes. The 617.3 nm line therefore experiences larger changes due to temperature effects than the 630.2 nm.

A direct comparison of plots shown in Fig. \ref{EQWvsTsim} and in Fig. \ref{EQWvsBsim} is not straightforward, as they show the variations of the line parameters with respect to two different physical quantities, the temperature and the magnetic field strength, respectively. Nevertheless, results from magnetic flux tube models show that within magnetic flux tubes in radiative equilibrium  the temperature at the formation heigths of the two investigated lines is positively correlated with the field strength, so that, for instance, model FAL-P can reasonably have associated a field strength of 1000-1500 Gauss. This allows a comparison of the trends of the line parameters with respect to the two physical quantities. The plots show that IC increases with the increase of both temperature and field strenght, in a manner difficult to disentangle. Opposite trends are instead found in the case of the FWHM, as an increase of temperature makes this line parameter to decrease while an increase of magnetic field strength makes it to increase. Comparison of the value of FWHM variation obtained for FAL-P with the value obtained for 1000 Gauss show that magnetic field effects are more likely to dominate. Variations of the EQW found in the case of the increase of the temperature are more than one order of magnitude larger than those obtained when increasing the magnetic field strength, again confirming that the direct effect of Zeeman splitting on this line parameter is negligible for the two lines.

Finally, it is important to notice that variations found from the numerical synthesis presented here are meant to qualitatively interpret observational results presented in Sec. \ref{measure}, as they are representative of the ideal case of unitary magnetic filling factor and observational effects like finite spatial and spectral resolution and scattered light have not been taken into account.

\section{Effects of unresolved magnetic features on observed lines}
\label{measurelowres}
 In this section we investigate whether and to what extent, line shape parameters are affected by small amounts of magnetic flux not resolved in the observations.

To this purpose we analysed average line shape parameters obtained in images that were
synthetized by replacing line profiles of pixels whose magnetic flux density exceeds a certain threshold with the average quiet-Sun profile. By varying the magnetic flux threshold,  images of successively lower average
magnetic flux density (in particular from approximately 90 to \textbf{10} Gauss) were constructed.  Note that the quiet Sun profile was computed averaging profiles of those pixels of the original images whose measured magnetic flux density is lower than 30 Gauss (see Sec. 2).  The line shape parameters were then determined from the average profiles of each of these synthetic maps, thus mimicing observations that do not resolve the remaining magnetic features. Figure \ref{shaveg} shows the relative differences between line shape properties from the synthetic frames (characterized by different values of the average magnetic flux) and those of the synthetic frame with the lowest magnetic flux density (i.e. 10 Gauss). In the case of IC and FWHM the error bars are derived from the uncertainties of the fitting of the line to a Gaussian function (see Sec. 2), while for the EQW the errors are estimated by the \citet{cayrel1988} formula.
We found that the average line shape parameters vary by a few percents with the increase of the average flux density. The maximum variation that we found are approximately $6\%$ in the case of IC for both lines, 4$\%$ and 2.5$\%$ in the case of FWHM for the 630.2 nm and the 617.3 nm, respectively, and -1.8$\%$ and -3.5$\%$ in the case of EQW.  Variations 
are larger in 617.3 nm in the case of EQW, as a result of the already mentioned larger sensitivity of this line to temperature. For the same reason, larger variations of FWHM with the magnetic flux density are instead found for the 630.2 nm. As explained above, in fact, for the 617.3 nm line the effect of temperature counteracting the Zeeman line broadening is larger, thus reducing the increase of the FWHM of this line with the flux.


\section{Discussion and Conclusions}
The measured line shape parameters of the two \ion{Fe}{1} lines vary with magnetic flux because of the direct effect of the Zeeman splitting and the indirect effect of temperature variations induced by suppression of convective motions.  IC and FWHM increase with the magnetic flux, while the opposite trend was found for EQW (Fig. \ref{shapes}). Results from numerical synthesis show that in the case of IC, the trends reflect both the Zeeman splitting and the temperature effects in a manner that is difficult to disentangle, whereas the FWHM is mostly sensitive to the Zeeman effect, in particular for larger field strenght. We have also shown that the dependence on magnetic field of the EQW is negligible at the values of magnetic flux investigated, so that the observed trend must be ascribed to temperature variations induced by the presence of magnetic field concentrations.

We also found that the three line shape parameters show saturation with the magnetic flux density (Fig. \ref{shapes}). Interpretation of these results in terms of magnetic flux tube models suggests that features characterized by small magnetic flux values are mostly spatially unresolved in our observations and that the pixels  characterized by the  largest magnetic flux values (above the saturation threshold) belong to spatially resolved features that may result from the clustering of magnetic elements. According to flux tube models, in fact, an increase of the magnetic field leads to an increase of the temperature within the tube in line-forming layers. Instead of a saturation  we should therefore observe IC and FWHM to monotonically increase and EQW to monotonically decrease with the flux. In particular, the saturation of EQW suggests that the magnetic features that we observed are mostly characterized by very similar photospheric temperatures. This is expected from models of clusters of magnetic flux tubes \citep[e.g.][]{criscuoli2009, fabiani1992}, and is in agreement with results recently reported by \citet{viticchie2010}. It is worth to notice that also \citet{harveylivingston1969a,harveylivingston1969b} concluded that the profiles of the \ion{Fe}{1} 525.0 and 525.3 nm lines suggest the "quantization" of the magnetic flux in the photosphere.

We also investigated the effects of unresolved magnetic field on the shapes of line profiles. Results reported in Fig. \ref{shaveg} show that variations of a few percent in the average shapes of the investigated lines are produced for modest (below 100 Gauss) values of average magnetic flux.  \textbf{Note that the average flux values investigated are compatible with those employed in irradiance reconstruction techniques to identify quiet Sun regions \citep[][and references therein]{fligge2000,ball2011}.} In the case of FWHM, larger variations with the average magnetic flux are found for the 630.2 nm than for the 617.3 nm, whereas in the case of EQW larger variations are found for the 617.3 nm. The 630.2 nm line is therefore more favorable to investigate magnetic field variation, while the 617.3 nm is more suitable to investigate effects of temperature variations.

It is interesting to notice that our estimates of EQW variations are in agreement with values reported by \citet{fabbian2010}
for the \ion{Fe}{1} 608.27 nm and 624.07 nm lines. These authors analyzed snapshots from Magneto Hydrodynamic Simulations to investigate the variations of average EQW values of some iron lines with the increase of the average magnetic flux in the simulations. Comparison of our Fig. \ref{shaveg} with their Fig. 4 reveals  that the agreement is better for the 617.3 nm line, as this has an excitation potential similar to that of 608.27 nm and 624.07 nm lines. Their plots also indicate that the variations of EQW mostly reflect variations of the average temperature in their simulations. The small dependence in their plots on the  average magnetic flux is due to the wavelenght shift of the opacity discussed in the Appendix. For completeness, we notice that those authors also reported a large dependence of EQW of the \ion{Fe}{1} 1.5 $\micron $ line on the magnetic flux. This is due on one hand to the fact that this line forms in regions where the variation of temperature induced by the magnetic field is negligible, on the other to its large Land\'{e} factor so that this line is not in the weak field regime.

The amount of variations of line shape parameters due to the presence of unresolved magnetic flux that we found is measurable with instruments like HINODE/SOT, SOLIS/VSM and SDO/HMI. Line shape variations of the investigated lines observed at disk center can be employed for studies of long term temporal variations of physical properties of quiet Sun. In particular, the results presented here show that it would be possible to disentangle thermal and magnetic effects by comparing  variations of different properties of the two \ion{Fe}{1} photospheric lines. For instance, a decrease over the magnetic cycle of EQW together with an increase of FWHM would indicate variations of quiet Sun temperature induced by the presence of unresolved magnetic features. On the other hand, a decrease  of the EQW together with a decrease of the FWHM would indicate an increase of the effective temperature of the plasma. 

Finally, it is worth to notice that \citet{martinezgonzalez2006} showed that the same profiles of Fe I 630.1 and 630.2 nm lines can be produced by different thermodynamic stratifications, magnetic field and microturbulence values, so that results obtained from spectro-polarimetric inversions should be treated with caution. However, the method that we propose is less affected by this degeneracy. In fact, we have shown that the EQW parameter of the two iron lines is almost insensitive to magnetic field effects, while the FWHM is mostly sensitive to magnetic field effects. The EQW variations can therefore be utilized to derive information on temperature variations, while the FWHM can be utilized to derive information on magnetic flux variations. Note, however, that a proper modeling of the observed data \citep[e.g.][]{holzreuter2012} is required to quantitatively estimate these variations. In particular, modifications of the velocity distributions, which can result from either variations of the magnetic flux or of the effective temperature, can also affect the shapes of lines. Nevertheless, since velocity distributions are known to narrow in the case of increase of magnetic flux \citep[e.g.][]{ishikawa2007,morinaga2008}, while they broaden and skew toward upflow velocities in the case of an increase of the effective temperature \citep[e.g.][]{gray1985,nordlund1990}, we expect velocity effects not to hamper the possbility of disentangling magnetic flux and effective temperature variations from observations of properties of the two lines.
We plan to investigate these effects in the future by analysing results from numerical simulations of the solar photosphere.

\acknowledgements
 This study was
supported by the Istituto Nazionale di Astrofisica (PRIN-INAF-2010) and the Agenzia Spaziale Italiana
(ASI/ESS/I/915/01510710). The NSO is operated by the Association of Universities for Research in
Astronomy, Inc. (AURA), for the National Science Foundation. IBIS was built by INAF/Osservatorio Astrofisico di Arcetri with
contributions from the Universities of Firenze and Roma ''Tor Vergata'',
the National Solar Observatory, and the Italian Ministries
of Research (MUR) and Foreign Affairs (MAE). We are grateful to A. Tritschler for providing the IBIS data reduction code.

\appendix 
\section{Dependence of Equivalent Width on magnetic field: analytical derivation}
\label{SecA1}
Here we investigate the dependence of the EQW of a magnetically sensitive spectral line on the magnetic field.
It is common practice to express the emergent line profile $I(\lambda)$ in the presence of Zeeman splitting as the sum of three Gaussians \citep[e.g.][]{borrero2008}:
\begin{equation}
I(\lambda)=1-\frac{(1-I_0)\sin^2\gamma}{2}e^{-\frac{(\lambda-\lambda _0)^2}{\Delta\lambda^2}} - \frac{(1-I_0)(\cos^2\gamma +1)}{4}\times ( e^{-\frac{(\lambda-\lambda _0-\lambda_B)^2}{\Delta\lambda^2}} + e^{-\frac{(\lambda-\lambda _0+\lambda_B)^2}{\Delta\lambda^2}} )
\label{eq1A}
\end{equation}
The two Gaussian functions shifted by the amount $\lambda_0 \pm \lambda_B$ represent the left- and right-hand circularly polarized $\sigma$ profiles, while the Gaussian function centered at $\lambda_0$ represents the profile of the linearly polarized $\pi$ component. $I_0$ and $\Delta\lambda$ are the core intensity and Doppler width of the line, in the absence of magnetic field; $\gamma$ is the inclination angle with respect to the line of sight; $\lambda_B = C g_{eff} \lambda_0^2 B$, where C is a constant, $g_{eff}$ is the effective Land\'e factor, $\lambda_0$ is the central wavelength of the line, and B is the magnetic field strength.

In the case of magnetic field oriented along the line of sight ($\gamma = 0 $) the previous formula simplifies to:
\begin{equation}
I(\lambda)=1 - \frac{(1-I_0)}{2}\times ( e^{-\frac{(\lambda-\lambda _0-\lambda_B)^2}{\Delta\lambda^2}} + e^{-\frac{(\lambda-\lambda _0+\lambda_B)^2}{\Delta\lambda^2}} )
\label{eq2A}
\end{equation}
The EQW of the line in that case is, by definition:
\begin{equation}
EQW:=\frac{1-I_0}{2}\bigr [\int_{0}^{\infty}e^{-\frac{(\lambda-\lambda _0-\lambda_B)^2}{\Delta\lambda^2}} d\lambda + \int_{0}^{\infty} e^{-\frac{(\lambda-\lambda _0+\lambda_B)^2}{\Delta\lambda^2}} d\lambda \bigl]
\label{eq3A}
\end{equation}
The two terms in the sum are equal as they differ only for by a shift within the infinite domain. Since the two terms represent the EQWs of the left ($EQW^{-}$) and right ($EQW^{+}$) circular polarized signal, then $ EQW = EQW^{-} = EQW^{+}$ (note that it is easy to verify that these equalities hold for any shape of the profiles, as far as the two circular polarized signals are symmetric).

We want now to study the variation of the EQW with the magnetic field. To this aim we differentiate Eq.\ref{eq3A} with respect to $B$:
\begin{equation}
\frac{d EQW}{dB} = \frac{1-I_0}{2} \frac{d}{dB}\Bigl[\int_{0}^{\infty}e^{-\frac{(\lambda-\lambda _0-\lambda_B)^2}{\Delta\lambda^2}} d\lambda + \int_{0}^{\infty} e^{-\frac{(\lambda-\lambda _0+\lambda_B)^2}{\Delta\lambda^2}} d\lambda \Bigr]
 \label{eq5A}
\end{equation}
Making the variable change $\tilde{\lambda} = \lambda - \lambda_0$, and changing the order of integration and 
differentiation, we obtain:
\begin{equation}
\frac{d EQW}{dB}=(1-I_0) Cg_{eff}\lambda_0^2
\Bigl [
\int_{-\infty}^{\infty}\frac{(\tilde{\lambda}-\lambda_B)}{\Delta\lambda^2}
e^{-\frac{(\tilde{\lambda}-\lambda_B)^2}{\Delta\lambda^2}} d\tilde{\lambda} +
\int_{-\infty}^{\infty}\frac{(\tilde{\lambda}+\lambda_B)}{\Delta\lambda^2}
e^{-\frac{(\tilde{\lambda}+\lambda_B)^2}{\Delta\lambda^2}} d\tilde{\lambda} \Bigr]
 \label{eq6A}
\end{equation}
where we have assumed that the rest wavelength is much larger then the Zeeman splitting. It is easy to see that the right-hand side of EQ. \ref{eq6A} is null, as the terms in
the integrand that are proportional to $\tilde{\lambda}$ are uneven, and the terms
that are proportional to $\lambda_B$ cancel. Thus, under the assumption that $I_0$ and $\Delta \lambda$ do not depend on the field, the EQW of a Zeeman split line is independent of the field strength. Howevere, this assumption does not hold in a stratified atmosphere, because the Zeeman splitting changes the opacity at a given wavelength, and therefore alters the height and physical parameters the line samples.
Indeed, as shown by previous studies \citep[e.g.][]{fabbian2010}, the EQW of a magnetic sensitive line is sensitive to the strength of the magnetic field.
These variations are small in the weak field case (i.e. $\lambda_0 \pm \lambda_B < \Delta\lambda$), and are non-negligible only for large fields. Moreover, the sensitivity of the EQW to the field depends on the atomic parameters of the line, as these determine its opacity.

\begin{figure} 
\epsscale{.9}
\plotone{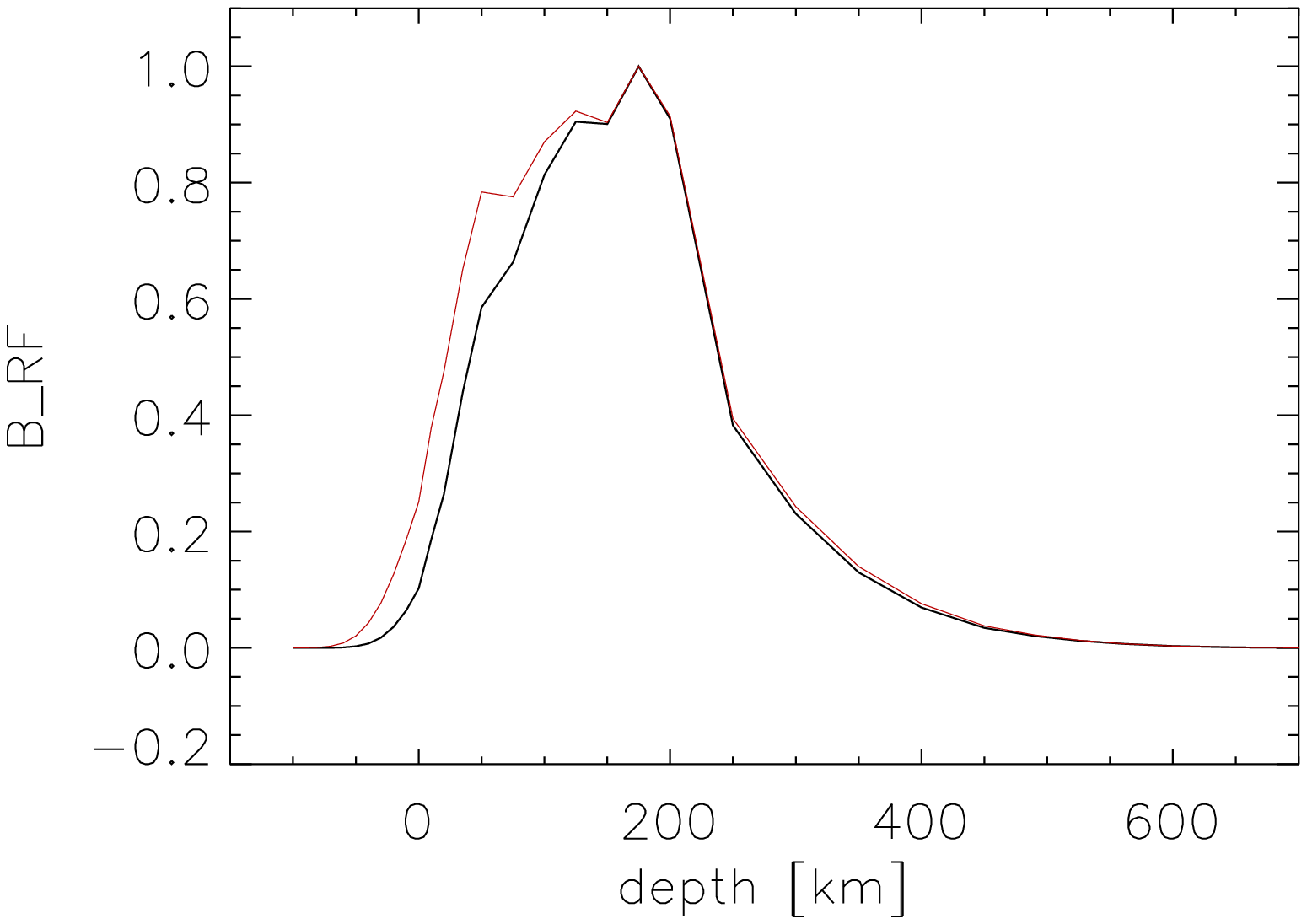}
\caption{Magnetic Field Response Functions of line core intensity obtained in FAL-P, representing facular regions, for the 630.2 nm (black) and the 617.3 nm (red) lines.
\label{response_B}  }
\end{figure} 

\begin{figure*} 
\epsscale{.4}
\plotone{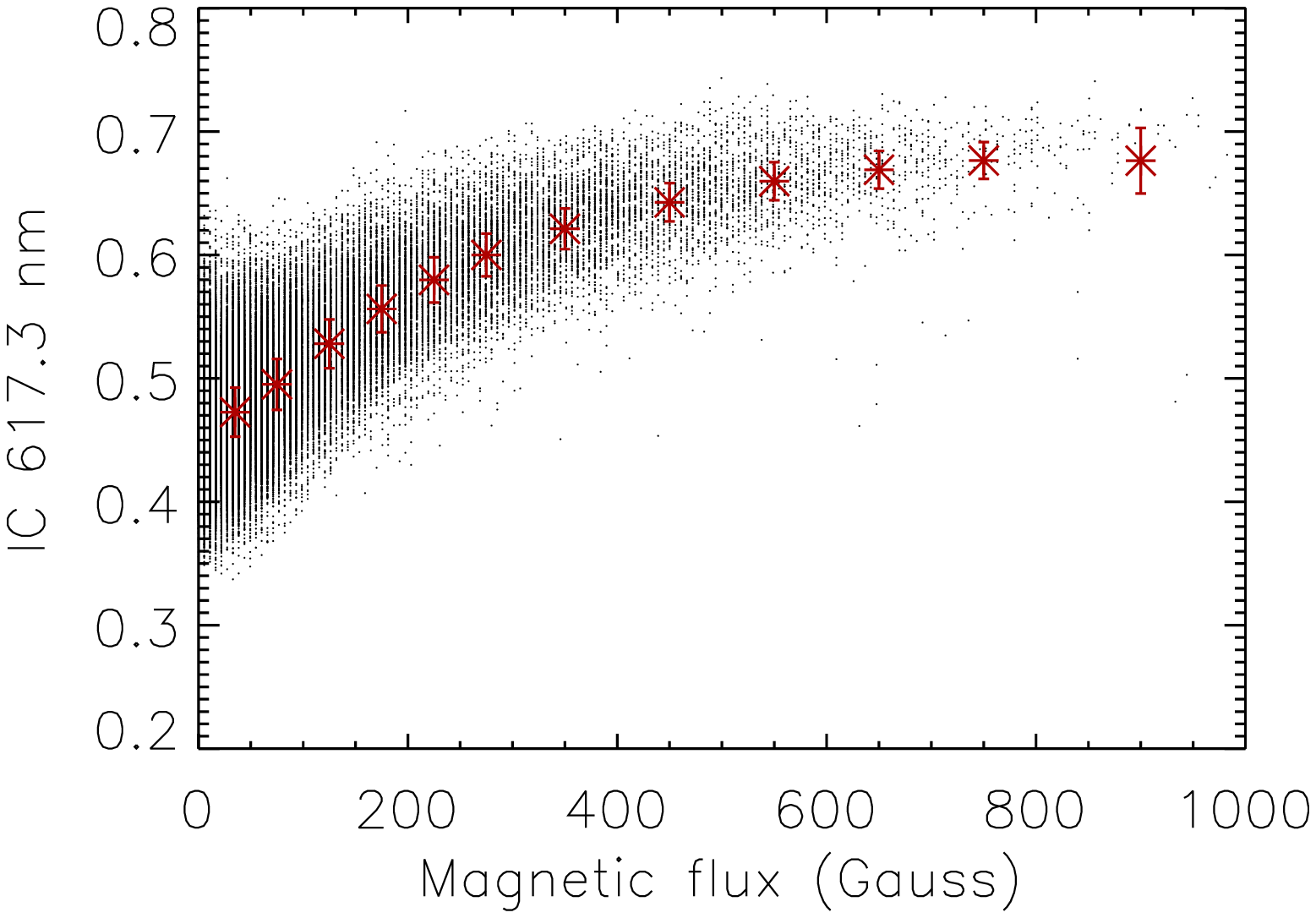}
\plotone{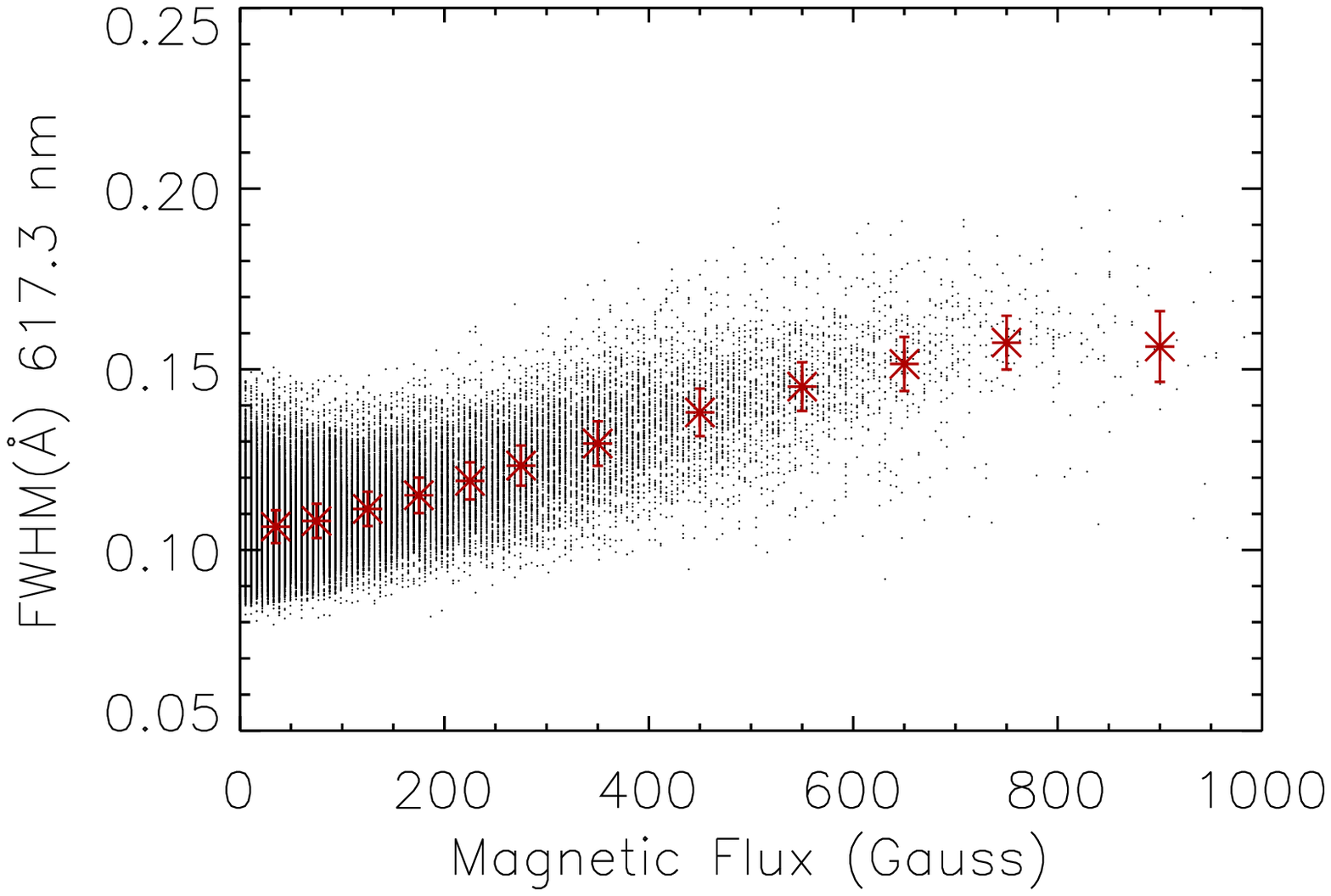}
\plotone{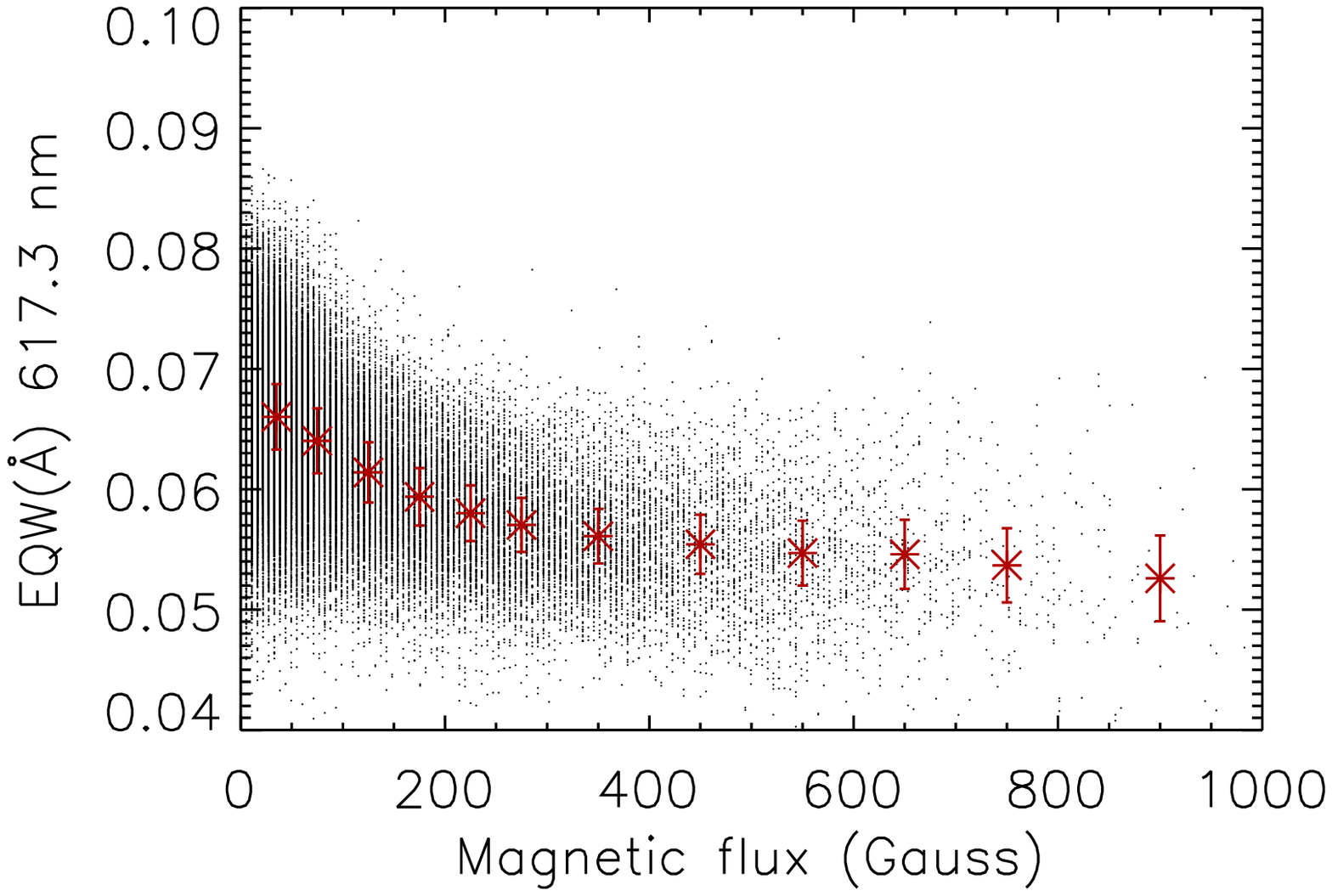}\\

\plotone{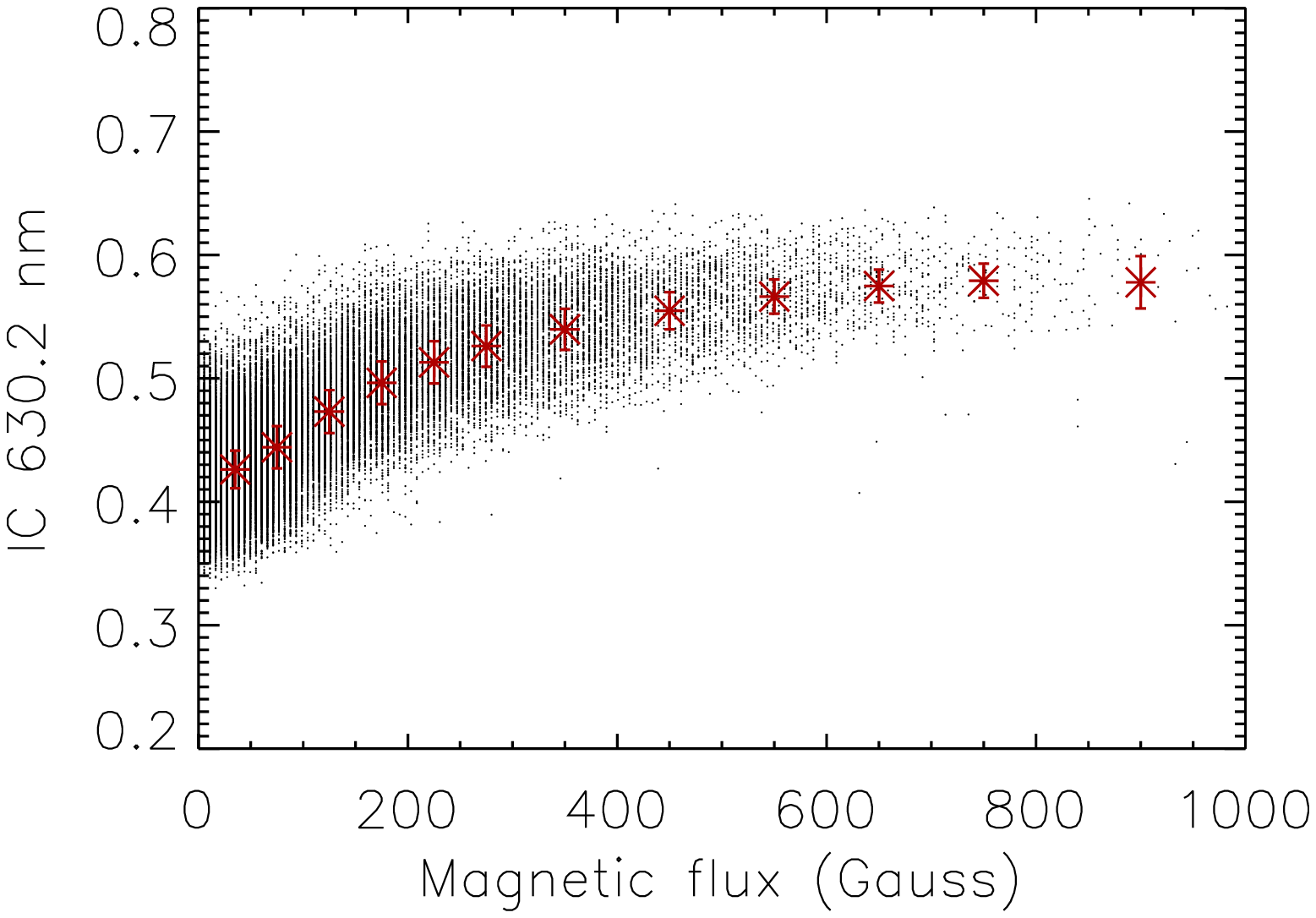}
\plotone{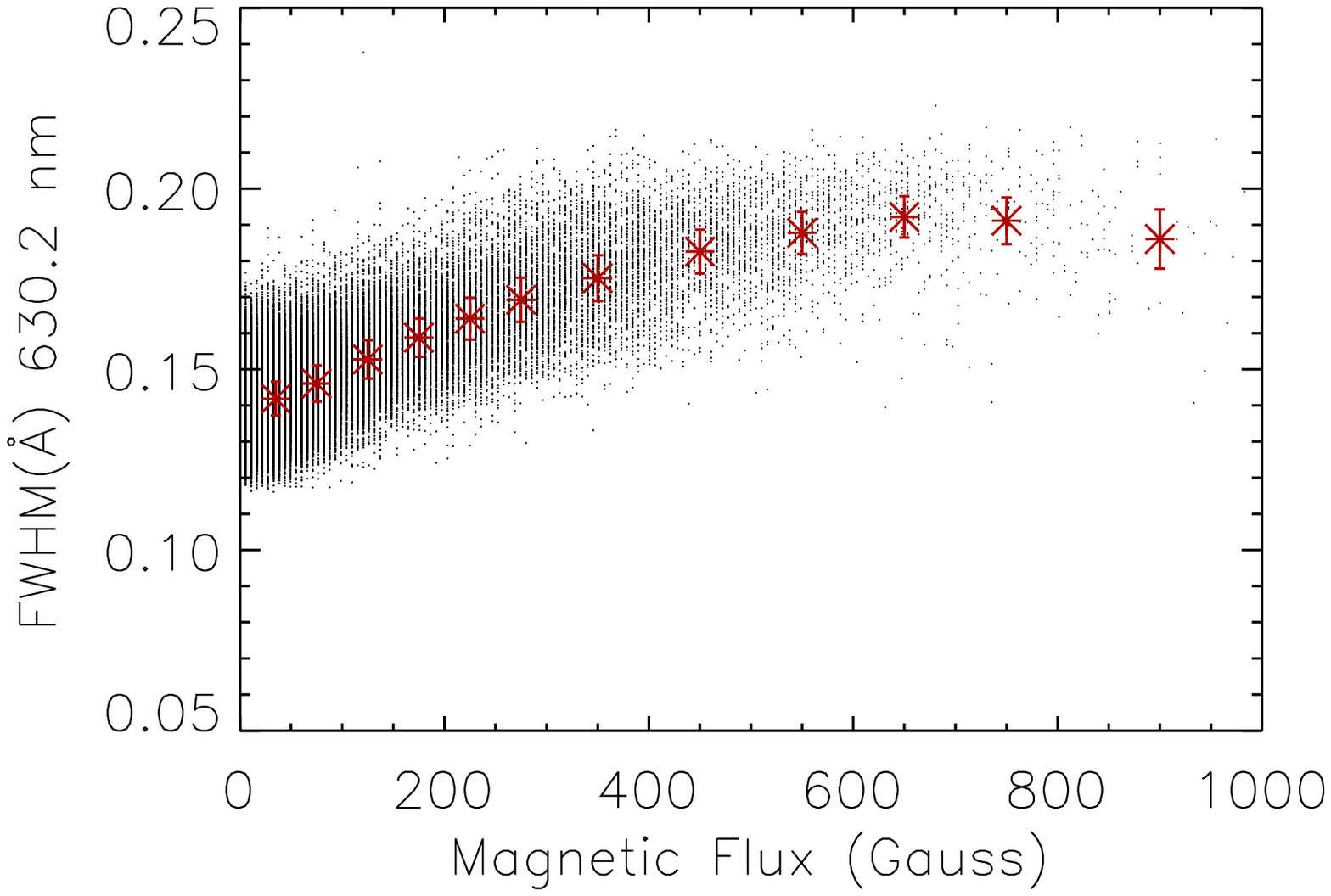}
\plotone{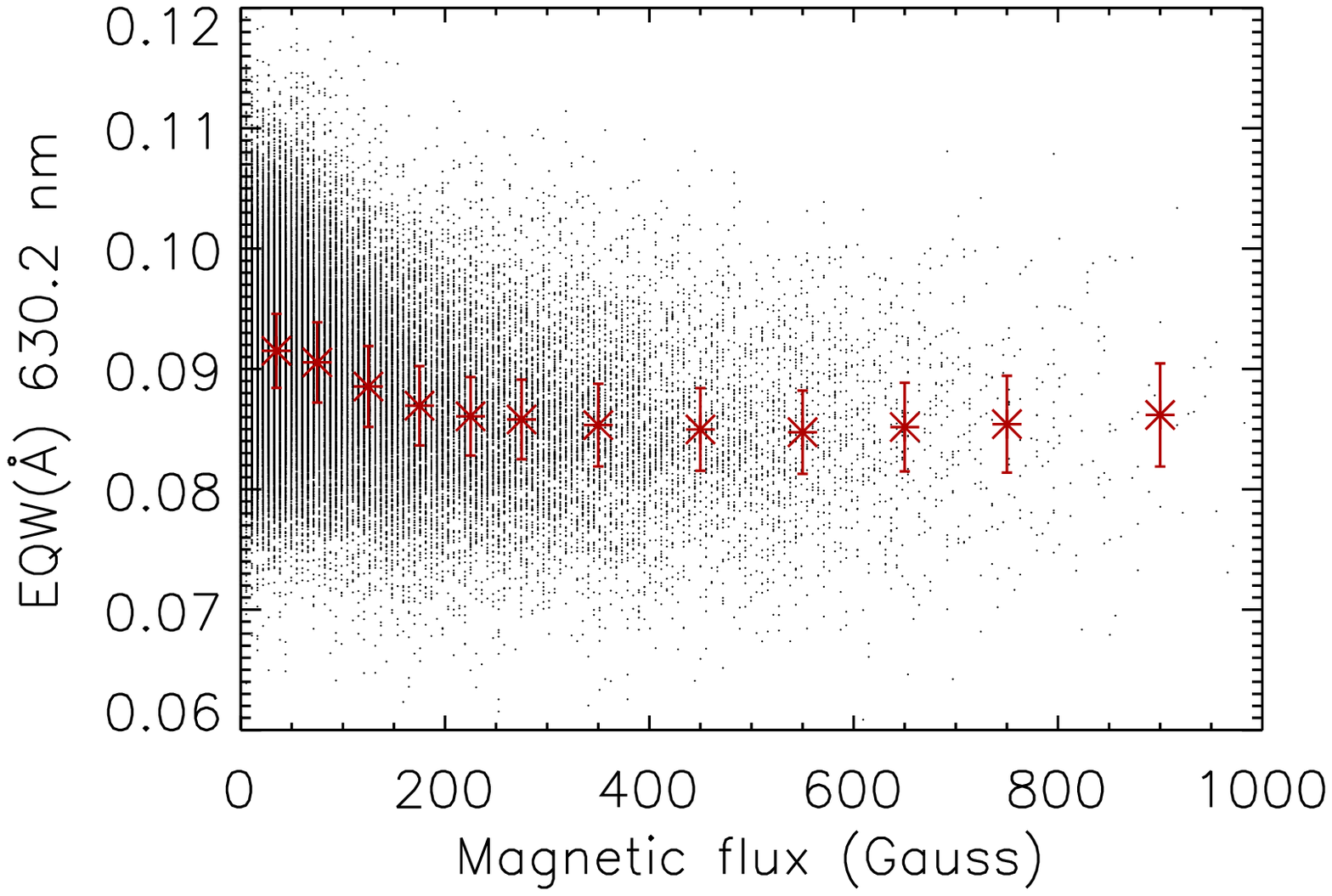}\\
\caption{Variations of line shapes parameters IC, FWHM and EQW for 617.3 nm (top) and 630.2 nm (bottom) lines with the magnetic flux density.
\label{shapes}  }

\end{figure*} 
\begin{figure*} 
\centering{
\epsscale{.8}
\plotone{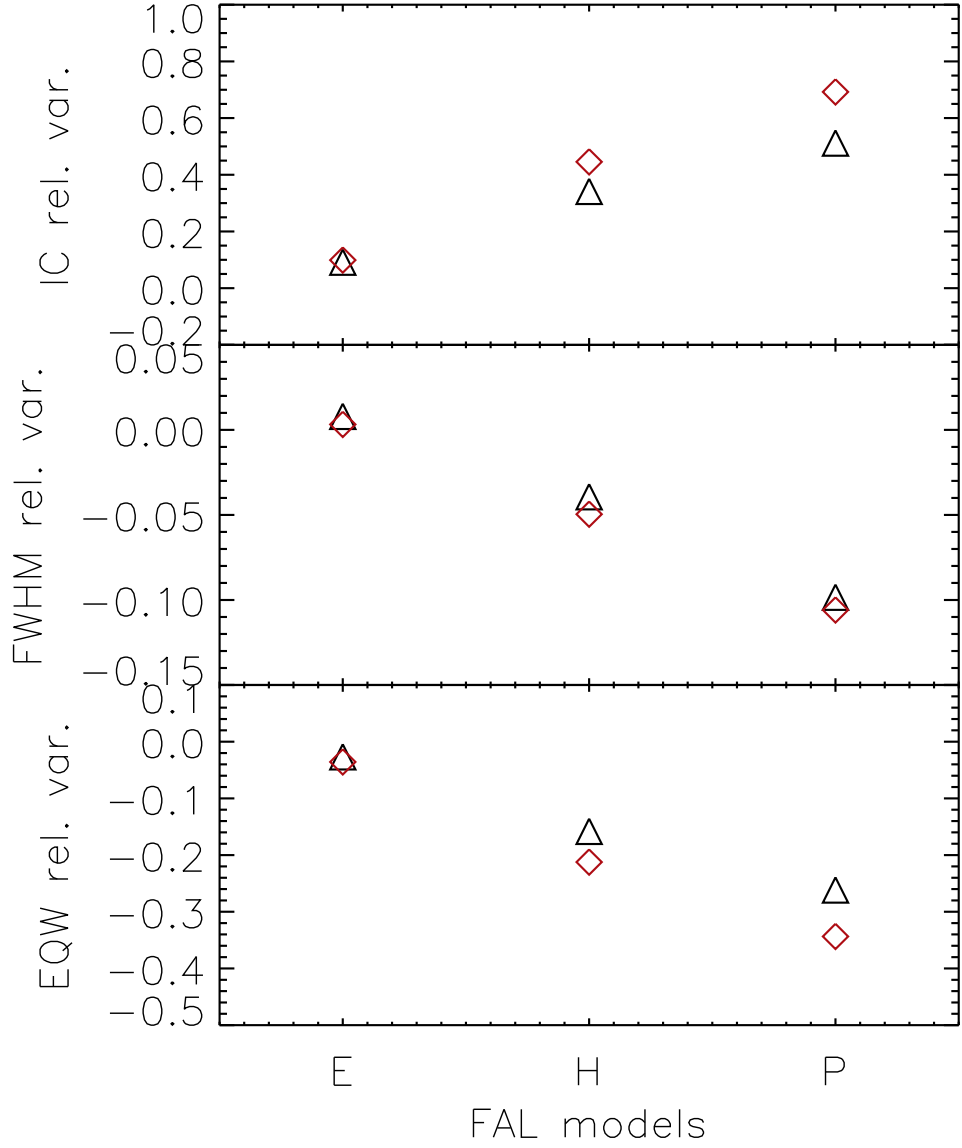}
\caption{Relative variation of line parameters obtained from spectral synthesis for different atmosphere models with respect to FAL-C.  Diamond: 617.3 nm; Triangles: 630.2 nm. \label{EQWvsTsim} }
}
 \end{figure*}

\begin{figure*} 
\centering{
\epsscale{.8}
\plotone{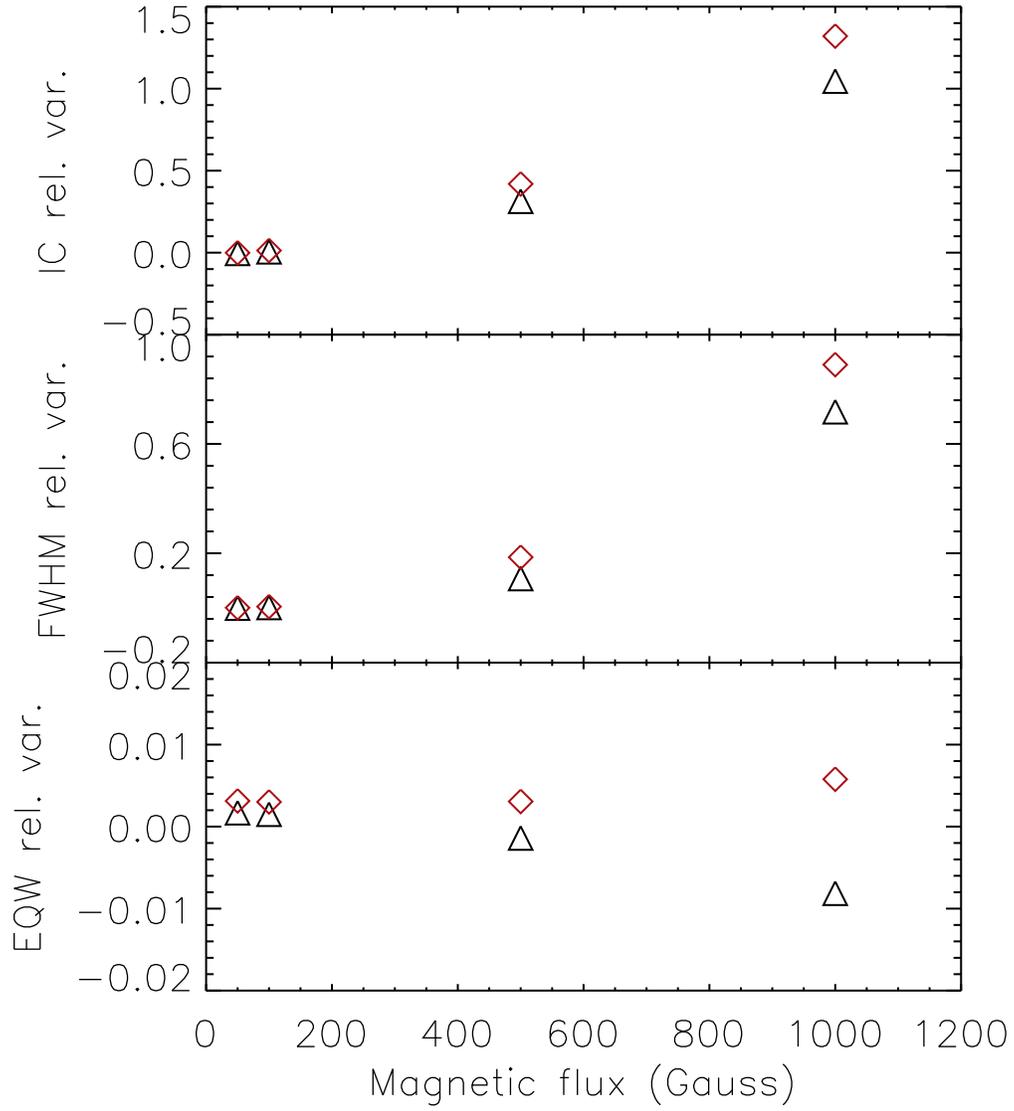}
\caption{Relative variation of line parameters  with the magnetic field strength obtained from spectral synthesis in FAL-C assuming uniform magnetic field stratification.  Diamond: 617.3 nm; Triangles: 630.2 nm. \label{EQWvsBsim} }
}
 \end{figure*} 


\begin{figure*} 
\centering{
\epsscale{.75}
\plotone{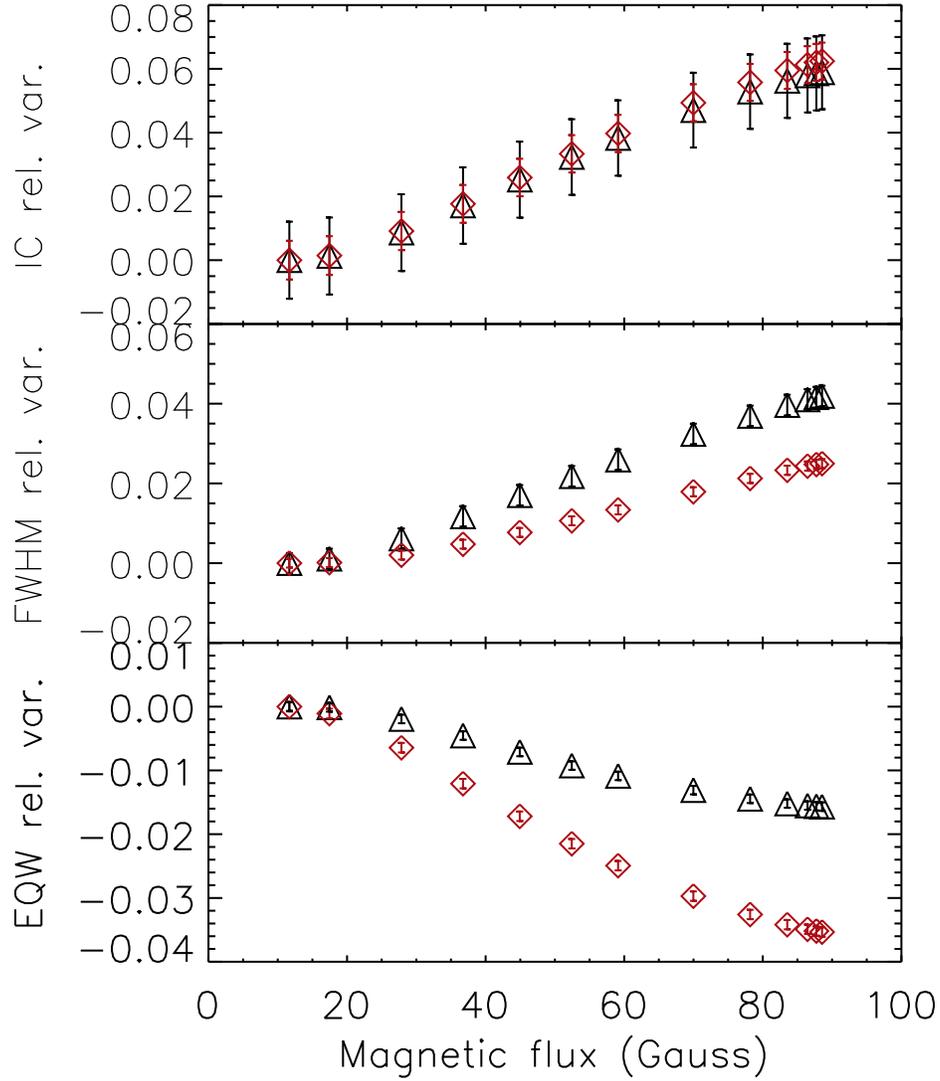}
\caption{Relative variation of average line shape parameters with the average magnetic flux density computed on synthetic frames (see text). Diamond: 617.3 nm; Triangles: 630.2 nm. \label{shaveg} }
}

\end{figure*}


\clearpage

\end{document}